\documentclass[11pt]{article}
\usepackage{amsmath}
\usepackage{amssymb}
\usepackage{graphics}
\def\cm{\checkmark}

\begin{document}

\title{Fragmentation of Spinning Branes}
\author{D.~Yamada
        \bigskip
        \\
       {\it Racah Institute of Physics},
       {\it The Hebrew University of Jerusalem},
        \smallskip
       \\
         {\it Givat Ram, Jerusalem, 91904 Israel}
          \bigskip
       \\
         {\tt daisuke@phys.huji.ac.il}}
\date{}
\maketitle

\begin{abstract}
The near-horizon geometries of spinning D3-, M2- and M5-branes
are examined by the probes immersed in a co-rotating frame.
It is found that the geometries become unstable at critical values
of the spin angular velocity by emitting the branes.
We show that this instability corresponds to the metastability of
the black hole systems and different from the known (local)
thermodynamic instability.
For the D3 case, the instability found here is in complete agreement
with the known metastability of
the $\mathcal{N}=4$ super-Yang-Mills theory with $R$-symmetry
chemical potentials.
\end{abstract}

\pagebreak

\tableofcontents
\pagebreak

\section{Introduction}

The main objective of this paper is to examine the behavior of
probes immersed in the near horizon geometries of spinning branes.
The motive originates from the AdS/CFT correspondence
\cite{Maldacena:1997re,Gubser:1998bc,Witten:1998qj}.
In its original and simplest form, the AdS/CFT correspondence
conjectures the equivalence between the $S^5$ compactification of
Type IIB supergravity and the strongly coupled $\mathcal{N}=4$
super-Yang-Mills theory in the 't Hooft limit.
This conjecture has been generalized to numerous varieties and one
of the most interesting cases is the correspondence between
the AdS black hole system and the finite temperature field theory
\cite{Witten:1998zw}.
One can further consider giving rotations to the black hole system.
One could give the rotations to the AdS bulk and consider the Kerr-AdS
black hole system \cite{Hawking:1998kw},
or to the $S^5$ and consider the $R$-charged black hole
system \cite{Chamblin:1999tk,Behrndt:1998jd,Cvetic:1999xp}.
Our current interest is the correspondence between
the latter black hole system in the grand canonical ensemble
and the finite temperature
$\mathcal{N}=4$ theory with the $R$-symmetry chemical potentials.

\begin{figure}[h]
{
\centerline{\scalebox{0.75}{\includegraphics{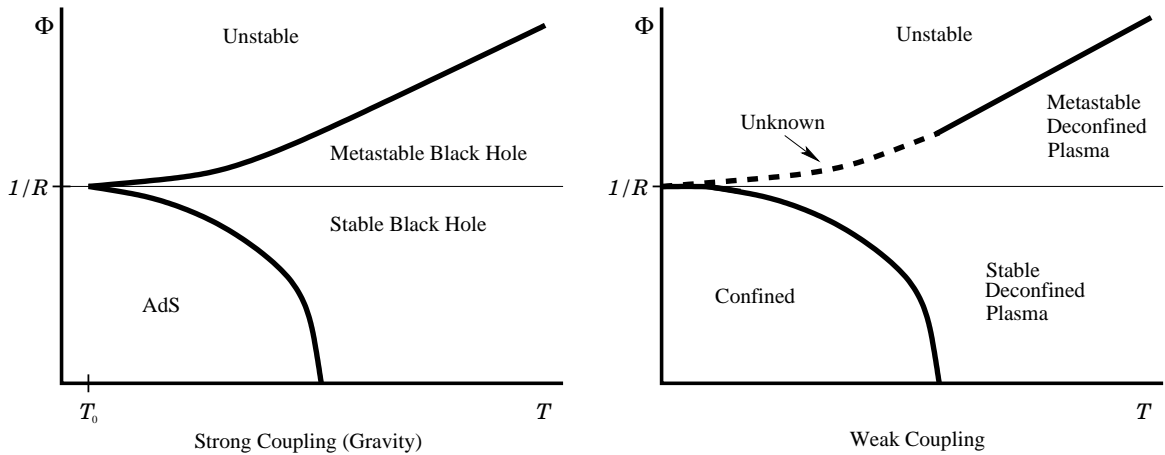}}}
\caption{\footnotesize
  The schematic drawing of the phase diagrams in the parameter
  space of the temperature $T$ and the chemical potential $\Phi$.
  The temperature $T_0$ in the gravity side is where the black hole
  horizon radius becomes zero.
  See Section~\ref{sec:M2M5} for more on this.
  The instability curve of the weak coupling analysis was obtained
  in the high temperature limit.
  The dotted curve is the extrapolation of the result
  to the low temperature region.
}\label{fig:comparison}%
}
\end{figure}

In Refs.~\cite{Chamblin:1999tk,Cvetic:1999ne,Yamada:2006rx,Yamada:2007gb},
the phase diagrams in the parameter space of the temperature
and the chemical potential are investigated for both sides
of the correspondence.
Despite the fact that the field theory analysis was carried out
in the weak coupling regime, the phase diagrams show impressive
qualitative agreement.
See Figure~\ref{fig:comparison}.%
\footnote{
  After this manuscript was sent to the journal, the paper by
  Hollowood {\it et al}. appeared \cite{Hollowood:2008gp}.
  They carried out the low temperature analysis of the weakly
  coupled field theory and obtained the result depicted as the
  dotted curve in the low temperature region.  
}
The Hawking-Page phase transition \cite{Hawking:1982dh} line
on the gravity side has the similar counter part in the field theory
side as the confinement/deconfinement phase transition line.
The local thermodynamic instability line of the black hole system
has the corresponding instability line of the plasma phase in
the field theory side.
There is one more known structure in the phase diagrams, namely,
the metastability lines of the systems.
The analysis of the black hole system
shows the existence of the metastability line \cite{Yamada:2007gb}.
If we denote the chemical potential of the system by $\Phi$,
the line is given as $\Phi=1/R$ with $R$ being the radius of the
AdS space, independent of the temperature.
In Ref.~\cite{Yamada:2006rx}, it is shown that
the field theory exhibits precisely
the same critical line at $\Phi=1/R$ in the phase diagram.%
\footnote{
  The field theory is defined on a three-sphere and $R$ is the radius
  of this sphere.
  Chronologically, the metastability line in the field theory
  phase diagram was found first.
}
It has been argued that the metastability of the field theory is
the manifestation of the violation of the BPS bound that follows
from the superconformal algebra.
Since the bound is protected against the quantum corrections and
the change in the coupling strength, it is pleasing that we have the
quantitative agreement between the metastability lines of the
black hole system and the finite temperature field theory.

In this paper, we would like to obtain further insight into
the metastability using the probe analysis.
There are two distinct hints that suggest the investigation using a probe.
First on the gravity side, it is known that the five dimensional $R$-charged
black hole solutions can be immersed into the $S^5$ compactification
of the ten dimensional supergravity theory \cite{Chamblin:1999tk,Cvetic:1999xp}.
Moreover, the ten dimensional geometry can be identified as
the near horizon limit of spinning D3-branes \cite{Cvetic:1999xp}.
Therefore, one can interpret the black hole as the effective
geometry created by the spinning D3-branes in the near horizon limit.
Second on the field theory side, the metastability was found by computing
the one-loop scalar effective potential.
The effective potential can be plotted against the expectation values
of various scalar eigenvalue configurations.
It was shown that the potential barrier that separates the metastable
state and the unstable direction is lowest when the potential profile
is made with respect to the expectation value of a single eigenvalue
splitting from the rest of the eigenvalues stack up in one place
\cite{Yamada:2006rx}.
Recall that if we view the $\mathcal{N}=4$ theory as the world
volume theory of the D3-branes, the scalar expectation value corresponds
to the location of the branes.
Then above two facts suggest that the metastability is caused by
the fragmentation of the branes from the spinning stack, and it
is most likely to occur by one-by-one splitting.
The probe analysis, therefore, is suitable for the investigation
of such phenomenon.
We can interpret the probe immersed in the background of the spinning branes
as a split brane from the stack and analyze the profile of the probe action
with respect to various locations in the bulk.
The analysis below confirms the expectation.

In the following sections, we carry out the probe analysis for the spinning
D3-branes, because this case has the concrete dual field
theory and we know its behavior.
Then in the appendices, we do the similar analysis for the
M2 and M5 cases.
The analyses start by establishing the higher dimensional
supergravity background geometry that we probe.
The geometries are asymptotically the product of the AdS space
and the sphere.
The sphere is ``twisted'' and the angular momentum and the angular
velocity correspond to the charge and the chemical potential, respectively,
as seen from the AdS bulk.
Then the probe is immersed in a co-rotating frame of the
background and the action is obtained.
The co-rotating frame is adopted to make the probe action
well-defined everywhere outside the horizon and to be
consistent with the conservation of the angular momentum.
Then the action is separately examined for the cases with flat and
spherical horizon geometries.
Finally apart from the probe analysis, we study
the reduced action of the corresponding lower dimensional
black hole system and show that the splitting
of the brane appears as the metastability of the black hole.

Along the analysis, it becomes clear that the one-by-one fragmentation
of the branes studied here
is different from the known (local) thermodynamic
instability of the spinning branes
\cite{Gubser:1998jb,Cai:1998ji,Cvetic:1999rb}.
The mechanism of the latter instability is not addressed in this paper
and is yet to be seen.

\section{Immersing Charged AdS$_5$ Black Holes
in Ten Dimensions}\label{sec:D3review}
We start with summarizing
the relevant results of Cveti\u c {\it et al.} \cite{Cvetic:1999xp}.
Along the way, we correct typos and establish our notations.
The idea of Ref. \cite{Cvetic:1999xp} is to obtain 
the charged AdS$_5$ black hole
solution of the five dimensional supergravity as the $S^5$ 
compactification of the ten dimensional one which can also
be identified as the near horizon limit of spinning D3-branes.

The action of the five dimensional $\mathcal{N}=2$ gauged
$U(1)^3$ supergravity can be written as \cite{Behrndt:1998jd}
\begin{align}\label{eq:S5}
  S_5 = &\frac{1}{16\pi G_5} \int d^5x\sqrt{-g}
        \bigg[ \mathcal{R} 
        - \frac{1}{2}\sum_{i=1}^3
               (\partial_{\mu} \ln X_i)(\partial^{\mu} \ln X_i)
        - \frac{1}{4}\sum_{i=1}^3 X_i^{-2}
                          F_{i\,\mu\nu}{F_i}^{\mu\nu}
        + \frac{V}{R^2} \bigg]
        \nonumber \\
        &+\frac{1}{16\pi G_5}
         \int F^{_{(2)}}_1\wedge F^{_{(2)}}_2\wedge A^{_{(1)}}_3
        \;,
\end{align}
where only the bosonic sector is shown.
The determinant of the metric is denoted by $g$
and $\mathcal{R}$ is the scalar curvature.
The fields $X_{i}$ with $i=1,2,3$ are the scalars
and they satisfy the constraint
\begin{equation}\label{eq:scalarConstraint}
  X_1X_2X_3 \equiv 1
        \;.
\end{equation}
The potential $V$ is defined as
\begin{equation}\label{eq:potential}
  V := 4 \sum_{i=1}^3 X_i^{-1}
        \;.
\end{equation}
The one-form fields $A^{_{(1)}}_i$ correspond to
the $U(1)^3$ gauged symmetry and $F^{_{(2)}}_i:=dA^{_{(1)}}_i$.
The parameter $R$ is the length scale
of the theory which is related to the cosmological
constant of the asymptotic AdS geometry.

The black solution is \cite{Behrndt:1998jd}
\begin{equation}\label{eq:d5soln}
  ds_5^2 = - H(r)^{-2/3} f(r) dt^2
        + H(r)^{1/3} \big[ f(r)^{-1}dr^2 + r^2 d\Omega_{3,k}^2 \big]
        \;,
\end{equation}
where the function $H(r)$ is
\begin{equation}
  H(r) = H_1(r)H_2(r)H_3(r)
          \quad
  \text{with}
          \quad
  H_i(r) = 1 + \frac{q_i}{r^2}
        \;,
\end{equation}
where, as we will see, the parameters $q_i$ are related to
the $U(1)^3$ charges.
We also have the function
\begin{equation}
  f(r) = k - \left(\frac{r_0}{r}\right)^2 + \left(\frac{r}{R}\right)^2H(r)
        \;,
\end{equation}
with three possibilities; $k=0,1$ or $-1$.
Those three cases have different horizon geometries, namely,
$\mathbb{R}^3$, $S^3$ and $\mathbb{H}^3$, respectively, and
the unit metrics $d\Omega_{3,k}^2$ are chosen accordingly.
In the following, we concentrate on the $k=0$ and $1$ cases,
because the former case is directly related to the near horizon
geometry of spinning D3-branes and the latter is to the field
theory defined on $S^3$.
The horizon is at $r=r_H$ which is the largest root of $f=0$.
The parameter $r_0$ is the ``non-extremality'' in that
$r_0=0$ corresponds to the extremal solution.
This parameter can be expressed in terms of others as
\begin{equation}\label{eq:r0}
  r_0 = r_H \bigg\{ k + \left(\frac{r_H}{R}\right)^2H(r_H) \bigg\}^{1/2}
        \;.
\end{equation}
The solutions to the scalars and gauge fields are
\begin{equation}\label{eq:Xi}
  X_i(r) = H(r)^{1/3}/H_i(r)
        \;,
\end{equation}
and
\begin{equation}\label{eq:Ai}
  A_{i\,\mu}(r) = -\frac{e_i}{r^2 + q_i}\delta_{\mu,0}
    \quad
    \text{with}
    \quad
    e_i:=\sqrt{q_i(r_0^2+kq_i)}
        \;,
\end{equation}
where $e_i$ are the $U(1)^3$ electric charges 
that satisfy Gauss's law.%
\footnote{
  The expression for $e_i$
  is valid for $k=0$ and $1$, and not for $k=-1$.
}

We would like to interpret the above five dimensional solution
as descending from a compactified higher dimensional theory.
Recall that the AdS$_5$ Schwarzschild black hole,
which is the special case of the above solution with
$k=1$ and $q_i=0$, is known to descend from
a solution to the $S^5$ compactification of Type IIB supergravity.
Therefore, the natural candidate for the higher dimensional
theory is the Type IIB compactified on $S^5$.
Also, the non-zero charges can be attributed to the momenta
in the extra compact space, so the rotations in $S^5$ are required.
Thus motivated, Cveti\u c {\it et al.} \cite{Cvetic:1999xp}
introduce the ansatz for the ten dimensional metric
\begin{equation}\label{eq:d10ansatz}
  ds_{10}^2= \tilde\Delta^{1/2}ds_5^2 +
                R^2\tilde\Delta^{-1/2}
                \sum_{i=1}^3X_i^{-1}
                \left\{ d\mu_i^2 + \mu_i^2 \big( d\phi_i + R^{-1}A_i^{_{(1)}} \big)^2
                  \right\}
  \;,
\end{equation}
where the function $\tilde\Delta$ is given by
\begin{equation}
  \tilde\Delta = \sum_{i=1}^3 X_i \mu_i^2
  \;,
\end{equation}
and the constraints (\ref{eq:scalarConstraint}) and
$\sum_i\mu_i^2=1$ are satisfied.
In the metric (\ref{eq:d10ansatz}), the part $ds_5^2$ is where we would like
the solution (\ref{eq:d5soln}) to fit in and the second term
describes the five-sphere with ``twists'' in $\phi_i$ directions.
The geometry must be supported by the self-dual five-form
field strength $(F^{_{(5)}} + *F^{_{(5)}})$ and the ansatz is%
\footnote{
  A sign is corrected from the original.
  This can be checked by comparing the equations of motion
  that follow from the Bianchi identity of the field strength
  and from the five dimensional action.
}
\begin{align}
  F^{_{(5)}} =& 
        (2/R)\sum_i\big(X_i^2\mu_i^2 - \tilde\Delta X_i\big)\epsilon^{_{(5)}}
        +(R/2) \sum_i (\bar{*}d \ln X_i) \wedge d(\mu_i^2)
        \nonumber \\
        &+(R^2/2) \sum_i X_i^{-2}d(\mu_i^2) \wedge
         \big( d\phi_i + R^{-1}A_i^{_{(1)}} \big) \wedge
         \bar{*}F_i^{_{(2)}}
  \;,
\end{align}
where $\epsilon^{_{(5)}}$ is the volume form with respect to the metric
$ds_5^2$ and $\bar{*}$ denotes the Hodge dual, also with respect to
the same five dimensional metric.

Plugging these ansatz into the Type IIB equations of motion,
one obtains the five dimensional equations of motion that can
be derived from the variation of the action (\ref{eq:S5}).
Thus the metric (\ref{eq:d5soln}), in fact, fits into
(\ref{eq:d10ansatz}).
The four-form $C^{_{(4)}}$ that satisfies
$F^{_{(5)}}=dC^{_{(4)}}$ is
\begin{equation}
  C^{_{(4)}} = 
        \bigg[
          \left( \frac{r}{R} \right)^4\Delta
          - \sum_i \frac{ r_0^2 + k(-r_H^2+q_i) }{R^2} \mu_i^2
        \bigg] dt \wedge \epsilon^{_{(3)}}
        +\sum_i\left(\frac{e_i}{R^2}\right)\mu_i^2
            (Rd\phi_i) \wedge \epsilon^{_{(3)}}
  \;,
\end{equation}
where $\Delta := H^{2/3}\tilde\Delta$, the volume form
$\epsilon^{_{(3)}}$ is with respect to $R^2d\Omega_{3,k}^2$
and $e_i$ are as defined in Eqn. (\ref{eq:Ai}).
We have corrected a sign of the above expression in the original
and included the constant $(r_0^2-kr_H^2)/R^2$ which does not
contribute to $F^{_{(5)}}$ but 
makes the value of
the probe action at $r=r_H$ zero.

Now, one of the points of Ref. \cite{Cvetic:1999xp} is that
the geometry (\ref{eq:d10ansatz}) with (\ref{eq:d5soln}) is
nothing but the effective geometry of spinning D3-branes
in the near horizon limit.
This is explicitly demonstrated in the reference for the $k=0$ case.
We can think of this case as the limit $r_H/R\to\infty$, {\it i.e.},
the large black hole limit of the $k=1$ case.
This allows us to think of the geometry as created by
the D3-branes and motivates us to the probe analysis
described in the introduction.

\section{Probing the Geometry of Post-Near-Horizon Limit}
\label{sec:D3post-NHL}

\subsection{Probe Action in Co-Rotating Frame}
\label{subsec:probeD3Action}

We now want to probe the geometry described in the previous section:
\begin{align}\label{eq:10dmetric}
  ds_{10}^2= &\tilde\Delta^{1/2} \big[
              - H^{-2/3} f dt^2
              + H^{1/3} \big\{ f^{-1}dr^2 + r^2 d\Omega_{3,k}^2 \big\}\big]
            \nonumber \\
             &+R^2\tilde\Delta^{-1/2}
                \sum_{i=1}^3X_i^{-1}
                \big\{ d\mu_i^2 + \mu_i^2 \big( d\phi_i 
                                + R^{-1}A_{i\,0}\,dt \big)^2
                  \big\}
  \;.
\end{align}
We note that this solution has the inverse temperature
$\beta = 4\pi (H^{1/2}/f')|_{r=r_H}$ and the angular velocities of the horizon
$\Omega_{H\,i}=-A_{i\,0}(r_H)/R$.%
\footnote{
  The temperature follows from the computation of the horizon
  surface gravity and the angular velocities can be obtained by
  requiring the co-rotating Killing vector field
  $\partial_t + \sum_i\Omega_{H\,i}\partial_{\phi_i}$
  to become null at the horizon.
}
Since we are interested in making the potential profile of 
the probe action with respect to the various locations of 
the probe, we consider the simple immersion map of
the probe D3-brane as shown in Table \ref{table:D3Immersion}.
\begin{table}[h]
  \centerline{
  \begin{tabular}{c|cccccccccc}
     & $t$ & $r$ & $y_1$ & $y_2$ & $y_3$ & $\theta$ & $\psi$ & $\phi_1$ & $\phi_2$ & $\phi_3$ \\ \hline
   D3 & \cm & & \cm & \cm & \cm & & & & &
  \end{tabular}
  }
  \caption{\footnotesize
    A representation of the immersion map of the D3-probe into the spacetime.
    The coordinates $y_i$ are in $d\Omega_{3,k}^2$ and $\theta$,$\psi$
    parametrize the directional cosines $\mu_i$.   
  }
\label{table:D3Immersion}
\end{table}
We do not allow the map to depend on the D3 time coordinate
so that the probe action does not contain kinetic terms,
desirable for us to make the profile.

This procedure, however, encounters immediate shortcomings.
Note that there is a region outside the horizon where the Killing
vector field $\partial_t$ becomes spacelike.
This is because while $f\to 0$ as $r\to r_H$, $A_{i\,0}$ approach
finite constants, so there is a range of $r$ where $g_{tt}>0$
in the metric (\ref{eq:10dmetric}).
This is the ergo-region that exists on the extra compact space $S^5$.
A probe inside this region must rotate (in the five-sphere) 
faster than the speed of light
to be static with respect to an asymptotically static observer.
The operational outcome is that the probe action becomes imaginary,
near but outside the horizon, because the action includes the square
root of the induced metric determinant.
Another problem is that we have $A_{i\,0}\to 0$ as $r\to\infty$.
This is unwanted because 
$A_{i\,0}(\infty)$ should be the chemical potentials of
the black hole system 
(see Refs. \cite{Braden:1990hw,Yamada:2007gb}),
as well as those of the boundary field
theory, so it is not consistent for these to vanish at the boundary.

There is a natural resolution that solves both problems at a stroke.
As emphasized by Hawking {\it et al.} in Ref. \cite{Hawking:1998kw},
there exists a co-rotating Killing vector field that is timelike
everywhere outside the horizon.
Though the reference deals with the Kerr-AdS black holes,
this fact is also true for our rotating $S^5$.
This is a special feature of asymptotically AdS spaces
and could never happen in the asymptotically flat case.
This suggests that we should immerse the probe brane in
a co-rotating frame of the background.
Now this procedure is equivalent to constructing the probe action
where the probe is rotating with respect to
an asymptotically static observer.
As mentioned in the introduction,
we interpret the probe as a split brane from the background-generating
spinning branes.
Then it is natural to consider the probe rotating such that
it conserves the angular momentum of the total system,
in particular at the horizon, the probe should be rotating
at the same velocity as the horizon.
We, therefore, move onto the co-rotating frame from which
the horizon appears static and the $S^5$ keeps finite
angular velocities at the boundary of the AdS bulk.

Such a frame can be obtained by a simple coordinate transformation
\begin{equation}
  \phi_i \to \phi_i + \Omega_{H\,i} \, t
  \;.
\end{equation}
It is then clear that the only change in the metric
(\ref{eq:10dmetric}) is the shift in $A_{i\,0}$,
\begin{equation}
  A_{i\,0} = \frac{e_i}{r_H^2+q_i} - \frac{e_i}{r^2+q_i}
  \;,
\end{equation}
and also the new form of $C^{_{(4)}}$ is given as
\begin{align}
  C^{_{(4)}} =& 
        \bigg[
          \left( \frac{r}{R} \right)^4\Delta
          + \sum_i \frac{1}{R^2}
            \bigg\{ \frac{e_i^2}{r_H^2+q_i}-(r_0^2 - kr_H^2+ kq_i) \bigg\} \mu_i^2
        \bigg] dt \wedge \epsilon^{_{(3)}}
        \nonumber \\
        &+\sum_i\left(\frac{e_i}{R^2}\right)\mu_i^2
            (Rd\phi_i) \wedge \epsilon^{_{(3)}}
  \;.
\end{align}
In the new frame, we have $A_{i\,0}\to 0$ as well as $f\to 0$ as
$r\to r_H$.
Consequently,
one can check that $g_{tt}$ of the metric is strictly negative
everywhere outside the horizon and this will make our probe
action well-defined for all $r\geq r_H$.
Moreover, the new $A_{i\,0}$ at the boundary have non-zero constants,
allowing us to interpret them as the chemical potentials of
the black hole system and of the boundary field theory.
Henceforth, we define 
\begin{equation}
  \Phi_i := A_{i\,0}(\infty)/R = \frac{1}{R}\frac{e_i}{r_H^2+q_i}
  \;,
\end{equation}
where the factor of $R$ is inserted for the dimensional reason,
and we call these the chemical potentials of the system.
Notice that the shift in $A_{i\,0}$ is a gauge transformation
as seen from the five dimensional theory described by the action
(\ref{eq:S5}).
And this is precisely the gauge chosen in the reduced action
analysis of Ref. \cite{Yamada:2007gb}.

In the following two subsections,
we are going to analyze the probe action immersed in the co-rotating
frame with the map represented in Table \ref{table:D3Immersion}.
The probe action has the form
\begin{equation}
  V_{D3} = WV - WZ
  \;,
\end{equation}
where we omit the overall minus sign and also drop
the volume factor, {\it i.e.},
the integral of $dt\wedge\epsilon^{_{(3)}}$,
and the tension of the brane.
The first term $WV$ represents the world volume part
with respect to the induced metric $G_{D3}$,
\begin{align}
  WV := &\sqrt{- \det G_{D3}}
        \nonumber \\
        =&\left( \frac{r}{R} \right)^3 \tilde\Delta
           \bigg( H^{1/3} f -
                  H \tilde\Delta^{-1}
                  \sum_i X_i^{-1}\mu_i^2 A_{i\,0}^2
           \bigg)^{1/2}
  \;,
\end{align}
and the second term $WZ$ represents the Wess-Zumino term,
that is, the RR-coupling part of the action. 
It is the pullback of the four-form
$C^{_{(4)}}$ with respect to the immersion map,
\begin{equation}
  WZ :=
        \left( \frac{r}{R} \right)^4 \Delta
          + \sum_i \frac{1}{R^2}
            \bigg\{ \frac{e_i^2}{r_H^2+q_i}-(r_0^2 -kr_H^2 + kq_i) \bigg\} \mu_i^2
  \;.
\end{equation}

\subsection{The Case with $k=0$}
\label{subsec:probeD3k0}
In this subsection, we analyze the probe action with $k=0$.
As mentioned earlier, the background geometry with $k=0$ arises
directly as the near horizon limit of spinning D3-branes and
the probe may be considered as a split brane from the spinning stack.
It is not too complicated to analyze the action in full generality
but for the sake of clarity,
we concentrate on representative special instances.
We, therefore, consider the cases with the charge configurations
$(q_1,q_2,q_3)=(q,0,0)$ and $(q,q,q)$.
In the following analyses, we set $R=1$ and measure the quantities
in the units of $R$.

\bigskip

\noindent {\it The $(q,0,0)$ Configuration.}
In this case, the probe action contains one of the two angles that
parametrize the directional cosines $\mu_i$.
It is easy to see that the potential is minimized when
the angle is $\pi/2$ and we concentrate on this special angle.
\begin{figure}[h]
{
\centerline{\scalebox{1.0}{\includegraphics{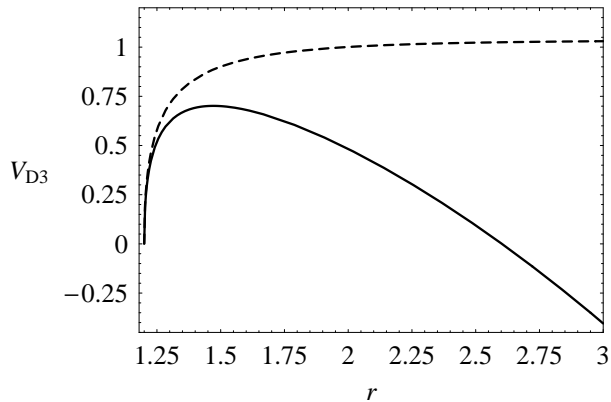}}}
\caption{\footnotesize
  The plot of the probe action in $(q,0,0)$ configuration.
  The position $r$ is measured in the units of $R$.
  The solid curve is plotted for $q=0.50$ and $r_H=1.20$
  which correspond to the chemical potential $\Phi=0.61$.
  The dashed curve is plotted for $q=0$ and $r_H=1.20$
  which correspond to the chemical potential $\Phi=0$.
}\label{fig:D3k0q}%
}
\end{figure}
Figure \ref{fig:D3k0q} is plotted for the chemical potential
$\Phi = 0.61$ in the solid curve
and for $\Phi = 0$ in the dashed curve.

In general, we have the instability for any $\Phi>0$ and
this can be seen in the asymptotic expansions
\begin{equation}
  WV = r^4 - \frac{1}{2} \Phi^2 r^2 + \mathcal{O}(1)
  \quad
  \text{and}
  \quad
  WZ = r^4 + \mathcal{O}(1)
  \;.
\end{equation}
This shows that when $\Phi=0$, the probe action approaches
constant at the asymptotic region due to the cancellation
between $WV$ attraction and $WZ$ repulsion.
Therefore, the probe ``feels'' no force acting on it at
the boundary.
This suggests that the configuration is in a BPS state,
and in fact, Bilal and Chu
showed that when a D3-brane is placed on the boundary of the (pure)
AdS$_5$ space, the configuration is fully BPS \cite{Bilal:1998ck}.
Therefore, we can conclude that the instability observed
for $\Phi>0$ is manifesting the violation of the BPS bound
at the boundary.

On the other hand, the behavior of the action near the horizon
is $WV\sim\mathcal{O}(\sqrt{r-r_H})$ and
$WZ\sim\mathcal{O}(r-r_H)$, so the Wess-Zumino part vanishes faster than
the world volume part, implying that the pull from the
world volume part wins out over the RR repulsion from $WZ$.
We can take the derivative of the potential with respect to
$r$ and examine the force that the probe feels near
the horizon:
\begin{equation}
  \frac{\partial V_{D3}}{\partial r} =
    r_H^3\sqrt{ r_H(r_H^2+q/2)/(r_H^2+q) } \; \frac{1}{\sqrt{r-r_H}}
    +\mathcal{O}(1)
  \;.
\end{equation}
This shows that the $WV$ pull always wins and
there exists no local instability near the horizon.
We will remark more on this point in Section~\ref{subsec:D3pre-NHL}.

\bigskip

\noindent {\it The $(q,q,q)$ Configuration.}
In this case, the probe action is independent of the two angles that
parametrize the directional cosines $\mu_i$.
This is because in this case, we have $\tilde\Delta = 1 = X_i$ and
the $S^5$ part of the metric (\ref{eq:10dmetric}) becomes
free of the $X$-scalar dependence.
\begin{figure}[h]
{
\centerline{\scalebox{1.0}{\includegraphics{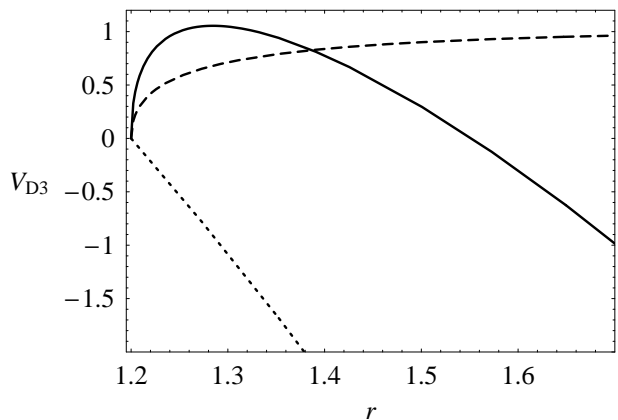}}}
\caption{\footnotesize
  The plot of the probe action in $(q,q,q)$ configuration.
  The position $r$ is measured in the units of $R$ and
  $r_H=1.20$ for all the cases.
  The solid curve is plotted for $q=2.00$
  which corresponds to the chemical potential $\Phi=2.19$.
  The dashed curve is plotted for $q=0$
  which corresponds to the chemical potential $\Phi=0$.
  The dotted curve is plotted for $q=2r_H^2$
  which corresponds to the chemical potential $\Phi=2.94$.
}\label{fig:D3k0qqq}%
}
\end{figure}
Figure \ref{fig:D3k0qqq} is plotted for the chemical potential
$\Phi = 2.19$ in the solid curve
and for $\Phi = 0$ in the dashed curve.
The asymptotic structure is similar to the case with $(q,0,0)$
configuration and we again have the instability for any $\Phi>0$.

The behavior near the horizon is slightly different.
We have, for this case,
\begin{equation}
  \frac{\partial V_{D3}}{\partial r} =
    \sqrt{ (r_H^2-q/2)(r_H^2+q)^3/r_H } \; \frac{1}{\sqrt{r-r_H}}
    +\mathcal{O}(1)
  \;.
\end{equation}
We thus see that there is a critical value of the charge
$q_c = 2r_H^2$ above which the probe action becomes ill-defined.
At the critical point, the attraction from the leading 
world volume part vanishes and the RR repulsion
causes the local instability at the horizon, unless $r_H=0$.
See the dotted curve of Figure \ref{fig:D3k0qqq}.
This behavior can be understood by noting
the temperature of the background
\begin{equation}
  \beta = \frac{\pi r_H^2}{(r_H^2-q/2)\sqrt{r_H^2+q}}
  \;.
\end{equation}
We see that at the critical value $q_c$, the temperature becomes zero
and beyond which, the system is not defined.%
\footnote{
  For the configuration $(q,0,0)$, we have
  $\beta=\pi\sqrt{r_H^2+q}/(r_H^2+q/2)$
  and the background temperature never becomes zero.
}
Therefore, the black brane (with $k=0$) at zero temperature
is locally unstable.

\bigskip

\noindent {\it To summarize}, we have found that any non-zero
chemical potential destabilizes the system against the one-by-one
fragmentation of the branes, if they tunnel through the potential
barrier that separates the near horizon and
the unstable asymptotic regions.
For any finite background temperature, the system is locally
stable against the brane splitting near the horizon.
Finally we remark that the configuration $(q,q,0)$ has the similar
behavior to the one with $(q,0,0)$.

\subsection{The Case with $k=1$}
\label{subsec:probeD3k1}

Let us now probe into the $k=1$ geometry which has the direct
relevance to the $\mathcal{N}=4$ SYM theory defined on $S^3$
and its phase structure \cite{Yamada:2006rx}.
As before, we are going to examine the representative configurations,
$(q_1,q_2,q_3)=(q,0,0)$ and $(q,q,q)$.

\bigskip

\noindent {\it The $(q,0,0)$ Configuration.}
In this case, just as in the situation with $k=0$,
the probe action contains one of the two angles that
parametrize the directional cosines $\mu_i$ and we
set this angle to $\pi/2$.
\begin{figure}[h]
{
\centerline{\scalebox{1.0}{\includegraphics{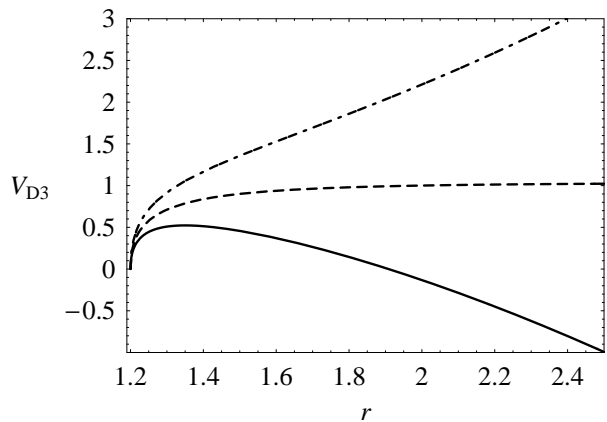}}}
\caption{\footnotesize
  The plot of the probe action in $(q,0,0)$ configuration.
  The position $r$ is measured in the units of $R$.
  For all the curves, $r_H=1.20$.
  The dash-dot curve is plotted for $q=0.10$
  which corresponds to the chemical potential $\Phi=0.40$.
  The dashed curve is plotted for $q=1$
  which corresponds to the chemical potential $\Phi=1$.
  The solid curve is plotted for $q=4.00$
  which corresponds to the chemical potential $\Phi=1.34$.
}\label{fig:D3k1q}%
}
\end{figure}
Figure \ref{fig:D3k1q} is plotted for the chemical potential,
$\Phi = 0.40$ in the dash-dot curve,
$\Phi = 1$ in the dashed curve
and for $\Phi = 1.34$ in the solid curve.

We have the critical value of the chemical potential
$\Phi_c=1$ above which the system becomes unstable.
This can be seen in the graphs as well as in the asymptotic
expansions
\begin{equation}
  WV = r^4 + \frac{1}{2}(1-\Phi^2) r^2 + \mathcal{O}(1)
  \quad
  \text{and}
  \quad
  WZ = r^4 + \mathcal{O}(1)
  \;.
\end{equation}
We observe that at the critical value $\Phi_c=1$, the probe on
the boundary
feels no force because the attraction from the world volume part
and the repulsion from the Wess-Zumino part exactly balance.
This is similar to the behavior of the $k=0$ case at $\Phi=0$.
Even though the analysis of Bilal and Chu \cite{Bilal:1998ck}
does not directly apply to the D3-probe of $k=1$ $S^3$ topology,
the analogy with the $k=0$ case
strongly suggests that we do have the BPS state at
$\Phi_c=1$ and the instability stems from
the violation of the BPS bound at the boundary.
We will comment more on this point, in connection with
the result from the dual field theory, later in the conclusion.

The probe action acts much similar way around the horizon as
in the $k=0$ case.
We have
\begin{equation}
  \frac{\partial V_{D3}}{\partial r} =
    r_H^3\sqrt{ r_H \{ r_H^2+(q+1)/2 \}/(r_H^2+q) } \; \frac{1}{\sqrt{r-r_H}}
    +\mathcal{O}(1)
  \;,
\end{equation}
showing that there exists no local instability near the horizon.

\bigskip

\noindent {\it The $(q,q,q)$ Configuration.}
In this case, just as in the situation with $k=0$,
the probe action is independent of the two angles that
parametrize the directional cosines $\mu_i$.
\begin{figure}[h]
{
\centerline{\scalebox{1.0}{\includegraphics{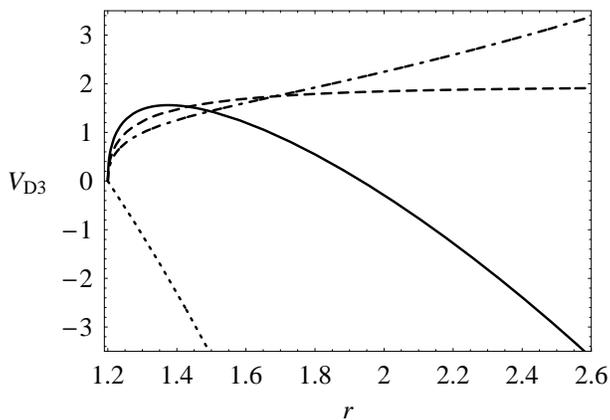}}}
\caption{\footnotesize
  The plot of the probe action in $(q,q,q)$ configuration.
  The position $r$ is measured in the units of $R$.
  For all the curves, $r_H=1.20$.
  The dash-dot curve is plotted for $q=0.10$
  which corresponds to the chemical potential $\Phi=0.41$.
  The dashed curve is plotted for $q=0.53$
  which corresponds to the chemical potential $\Phi=1.00$.
  The solid curve is plotted for $q=1.50$
  which corresponds to the chemical potential $\Phi=1.89$.
  The dotted curve is plotted for $q=2.99$
  which corresponds to the chemical potential $\Phi=3.14$.
}\label{fig:D3k1qqq}%
}
\end{figure}
Figure~\ref{fig:D3k1qqq} is plotted for the chemical potential,
$\Phi = 0.41$ in the dash-dot curve,
$\Phi = 1.00$ in the dashed curve,
$\Phi = 1.89$ in the solid curve
and $\Phi = 3.14$ in the dotted curve.
The asymptotic structure is similar to the case with $(q,0,0)$
configuration and we again have the instability for $\Phi>1$.

The behavior near the horizon is slightly different from $(q,0,0)$
but similar to $k=0$.
We have, for this case,
\begin{equation}
  \frac{\partial V_{D3}}{\partial r} =
    \sqrt{ \{2r_H^6+(1+3q)r_H^4-q^3\}(r_H^2+q)/(2r_H) } \; \frac{1}{\sqrt{r-r_H}}
    +\mathcal{O}(1)
  \;.
\end{equation}
We thus see that there is a critical value of the charge, $q_c$,
satisfying $2r_H^6+(1+3q_c)r_H^4-q_c^3=0$,%
\footnote{
  There is only one real root to the equation and we take that
  to be $q_c$.
}
above which the probe action becomes ill-defined.
At the critical point, the attraction from the leading 
world volume part vanishes and the RR repulsion of WZ-term
causes the local instability at the horizon, unless $r_H=0$.
See the dotted curve of Figure \ref{fig:D3k1qqq}.
As one can see in the background temperature
\begin{equation}
  \beta = \frac{2\pi r_H^2(r_H^2+q)^{3/2}}{2r_H^6+(1+3q)r_H^4-q^3}
  \;,
\end{equation}
the critical value of the charge corresponds to zero temperature
and beyond which the system is not defined.%
\footnote{
  The inverse temperature for $(q,0,0)$ is
  $\beta = 2\pi\sqrt{r_H^2+q}/(2r_H^2+q+1)$
  and the background temperature never becomes zero.
}
We see that the black hole system at zero temperature
is locally unstable.

\bigskip

\noindent {\it To summarize}, we have found the critical value of
the chemical potential, $\Phi_c=1$,
above which the system becomes unstable against the one-by-one
fragmentation of the branes, if they tunnel through the potential
barrier.
For any finite background temperature, the system is locally
stable against the brane splitting near the horizon.
Finally we remark that the configuration $(q,q,0)$ has the similar
behavior to the one with $(q,0,0)$.

\subsection{Comments on the Geometry of Pre-Near-Horizon Limit}
\label{subsec:D3pre-NHL}

We have so far examined the effective geometry of spinning
D3-branes in the near horizon limit.
We now briefly comment on the geometry of the branes
without taking the limit.
This geometry, being asymptotically flat, has no Killing vector field
that is timelike everywhere outside the horizon,
and the connection to the dual field theory is not direct.
However, we have interpreted the probe brane
as a brane splitting from the stack of spinning branes
and such picture probably is more intuitively straightforward
in the geometry before taking the near horizon limit.
The main reason for probing this geometry is to make sure
that the lack of the local instability near the horizon observed
in the previous subsections is not due to the limit and show that
the local thermodynamic instability of the spinning
branes found in Refs.~\cite{Gubser:1998jb,Cai:1998ji,Cvetic:1999rb}
does {\it not} correspond to the one-by-one fragmentation
of the branes.
This geometry should suffice for this purpose, though the behavior
of the probe far away from the horizon is unclear.

The asymptotically flat geometry of spinning D3-branes is shown
in Refs. \cite{Cvetic:1999xp,Kraus:1998hv,Russo:1998by}.
One can compute the angular velocity of the horizon and change
the metric to the co-rotating frame in which the horizon
appears static.
The same transformation should be applied to the RR four-form
of the theory as well.
Then the probe action can be obtained with respect to the similar
immersion map as represented in Table \ref{table:D3Immersion}.
This action is ill-defined far away from the horizon due to
the presence of the ergo-region, but behaves properly near the horizon.
Restricting the analysis to the region near the horizon, we find
the action behaving very similar to the previous post-near-horizon-limit
analyses; we have no local instability for any finite temperature.
This shows that the special local instabilities of the previous
subsections are
different from the thermodynamic instability of the spinning branes
which is present at finite temperature.

\section{Analysis of Reduced Action}
\label{sec:reducedD3Action}

In this section, we show that the observed
brane fragmentation corresponds to the metastability
of the five dimensional black hole (or brane) system.
Since the metastability is not a local phenomenon,
the usual action evaluated
on a known solution -- the on-shell action -- is not 
appropriate for our purpose.
We need the action, known as the reduced action
\cite{Whiting:1988qr,Braden:1990hw},
that is valid off shell in the space of physical parameters.
The construction of the reduced action for the system of
our interest is detailed in Ref. \cite{Yamada:2007gb}.
Though the reference deals only with the $k=1$ case,
the extension to the $k=0$ case is straightforward.
We thus refer the reader to the paper for the details
and here we merely describe the idea of the construction.

The construction
starts with the Euclidean action of the system (including the Gibbons-Hawking
and counter terms) and one places it in a finite box.
Then, restrict the analysis to a class of geometries
that satisfy certain conditions, such as static spherical symmetry.
The restriction results in a specific form of the metric and
this is analogous to a metric ansatz that one makes in solving
the Einstein equations.
In addition to the geometrical restriction, one imposes
the appropriate Euclidean black hole
topology on the manifold, yielding
a condition that the metric must satisfy.
Next, the thermodynamic data is encoded on the wall of the box.
The data, in the case of a grand canonical ensemble, are the
temperature and chemical potential and they are encoded by
setting the circumference of the $S^1$ time circle and the value of
the gauge field, respectively, at the wall.
Then, the Hamiltonian constraint and Gauss' law are solved and
eliminated from the action functional and it is integrated
with the topology condition and the boundary data.
Finally the wall of the box is removed to infinity while rescaling
the temperature and the chemical potential appropriately,
resulting in the simplified reduced action
parametrized in terms of the horizon radius and the electric charge.
Notice that the dynamical equations of motion have not been used
in reducing the action.
Thus the reduced action is valid throughout
the parameter space and a black solution corresponds to
a saddle point of the action in the parameter space.
Therefore, this action
captures the global structure of the theory.

The resulting action has the form
\begin{equation}
  I_* = \beta \bigg( E - \sum_i e_i\Phi_i \bigg) - S
  \;,
\end{equation}
where
\begin{equation}
  E := \frac{1}{R^3} \bigg( \frac{3}{2}r_0^2 + k \sum_i q_i \bigg)
  \;,
  \quad
  S := 2\pi \left( \frac{r_H}{R} \right)^3 H(r_H)^{1/2}
  \;,
\end{equation}
and the parameters $e_i$, $r_0$ and the function $H$ are as defined
in the previous sections.
The parameters $\beta$ and $\Phi_i$ are independent input data,
as the grand canonical ensemble should be.
We note that the overall dimensionless constant $\omega_3/8\pi G_5$,
with $\omega_3$ being the integral of $\epsilon^{_{(3)}}$, is
omitted and the Casimir energy which appears for the $k=1$ case
is also dropped.
Extremized with respect to the parameters
$r_H$ and $q_i$, the reduced action correctly reproduces the on-shell
expressions for $\beta$ and $\Phi_i$.
Moreover, considering $I_*$ as the zero-loop approximation to
the free energy of the system, one can compute the thermodynamic
quantities by taking the appropriate derivatives of $I_*$.
Then it is easy to verify that the resulting quantities satisfy
the first law of thermodynamics when evaluated at the on-shell
values of $\beta$ and $\Phi_i$.

The stability of the system for given temperature and chemical potentials
can be observed from the behavior of
the action with respect to the parameters $r_H$ and $q_i$.
In particular, the local stability can be examined by computing
the Hessian of the action with respect to $r_H$ and $q_i$,
as was done for the $k=1$ $(q,0,0)$ case by
Cveti\u c and Gubser in Ref. \cite{Cvetic:1999ne}.

The results are summarized in the following two subsections.
We will find that the metastabilities of the systems set in
exactly at the same critical values of the chemical potentials
as the probe analysis.
The local thermodynamic instabilities are also found but they are
different from the special local instabilities of the probe analysis.

\subsection{The Case with $k=0$}
\label{subsec:reducedD3k0}

\begin{figure}[h]
{
\centerline{\scalebox{0.8}{\includegraphics{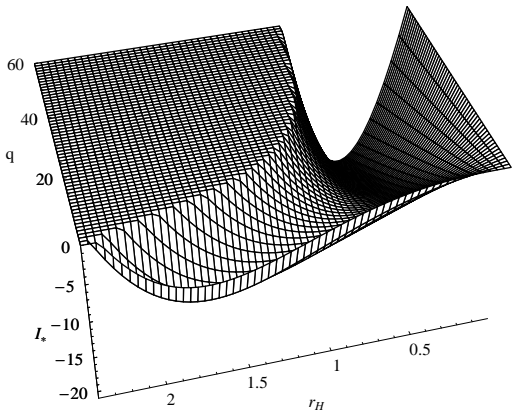}}}
\caption{\footnotesize
  The plot of $I_*$ against $r_H$ and $q$ for $k=0$ and $(q,0,0)$ case.
  The temperature and chemical potential are set to
  $T=0.55$ and $\Phi = 0.9$, both measured in the units of $R$.
  The local minimum is at $(r_H,q)\approx(1.71,1.12)$ and
  the unstable saddle point is at $(0.94,9.20)$.
}\label{fig:D3Redk0}%
}
\end{figure}

Figure \ref{fig:D3Redk0} is the plot of the reduced action
for the case with $k=0$ in $(q,0,0)$ configuration.
In the plot, we see the existence of the local minimum, corresponding
to the large black brane solution, and the unstable direction,
separated by the unstable saddle point corresponding to the small
black brane solution.
In other words, the large black brane is metastable.
The expansion of the reduced action,
\begin{equation}
  I_* = \beta r_H\big( 3r_H/2 - \Phi \big)q + \mathcal{O}(\sqrt{q})
  \;,
\end{equation}
shows that the unstable direction exists for any $\Phi>0$
in the direction of large $q$ and small $r_H$.
This is in agreement with the probe analysis.

In the configuration $(q,q,0)$ or $(q,q,q)$, {\it i.e.},
when we have the constraint $q_1\equiv q_2$ or
$q_1\equiv q_2\equiv q_3$, we do not have the metastability.
This is because the constraints reduce the dimension of the parameter
space and shut the unstable direction.
However, notice that these configurations are continuously connected
to the $(q,0,0)$ case in the parameter space.
Therefore, if we allow the parameters to fluctuate around the constraints,
nothing prevents the system to decay into the unstable direction
found above.
We thus have the instability for generic configurations with
$\Phi_i>0$.

The local instability has the similar behavior.
One finds the instability for $(q,0,0)$ but not
under the constraints $q_1\equiv q_2$ and $q_1\equiv q_2 \equiv q_3$.
Allowing the constraints to fluctuate, one discovers the local
instabilities similar to the $(q,0,0)$ case.
This means that the surfaces of the constraint equations are
locally unstable in the larger parameter space.
The thresholds of the stability appear as straight lines with
zero intercepts in the $T$-$\Phi$
plane and the slopes of the lines are
listed in the table below for some cases.
\begin{table}[h]
  \centerline{
  \begin{tabular}{c|ccc}
    $(q_1,q_2,q_3)$  &   $(q,0,0)$  &  $(q,q,0)$  &  $(q,q,q)$  \\ \hline
    Slope     &$\pi/\sqrt{2}$&  $\pi$      &  $2\pi$
  \end{tabular}
  }
\end{table}
(The temperature and chemical potential are taken as the horizontal
and vertical axes, respectively. See Figure~\ref{fig:D3k1Local} for
the actual plot.)
At these critical lines, the large and small black brane saddle points
merge and this opens up the unstable direction without the potential
barrier.
When the chemical potential becomes larger than this critical value
at a fixed temperature, the saddle point of the action cease to exist.%
\footnote{
  One can also obtain the critical line in $q$-$r_H$ plane similar
  to the one in Ref. \cite{Cvetic:1999ne}.
  In this case, crossing the critical line corresponds to moving
  from large to small black brane solution at a given set of the
  temperature and the chemical potential.
}

Recall that the local instability of the probe action in the $(q,q,q)$
configuration occurred only at zero temperature.
The local instability of the reduced action observed here is, therefore,
essentially different.

\subsection{The Case with $k=1$}
\label{subsec:reducedD3k1}

\begin{figure}[h]
{
\centerline{\scalebox{0.9}{\includegraphics{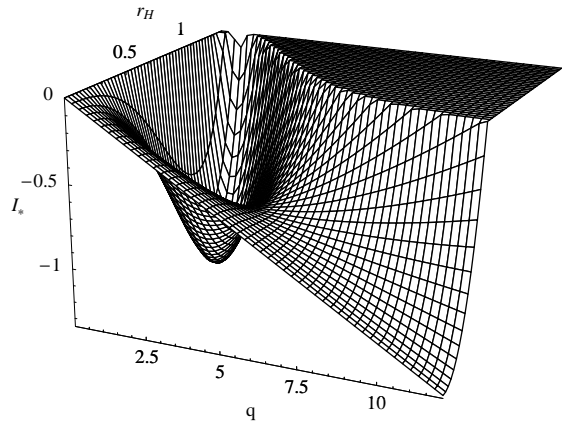}}}
\caption{\footnotesize
  The plot of $I_*$ against $r_H$ and $q$ for the $k=1$ $(q,0,0)$ case.
  The temperature and chemical potential are set to
  $T=0.45$ and $\Phi = 1.05$, both measured in the units of $R$.
  The local minimum is at $(r_H,q)\approx(0.996,1.23)$ and
  the unstable saddle point is at $(0.359,5.38)$.
}\label{fig:D3Redk1}%
}
\end{figure}

This case has been studied in detail in Ref. \cite{Yamada:2007gb}.
The plot of the reduced action for the $(q,0,0)$ configuration
is reproduced in Figure \ref{fig:D3Redk1}.
This case has the critical metastability line at $\Phi_c=1$ independent
of the temperature, in agreement with the probe analysis.
The other parameter configurations behave similar to what have been
described in the $k=0$ case.

\begin{figure}[h]
{
\centerline{\scalebox{1.2}{\includegraphics{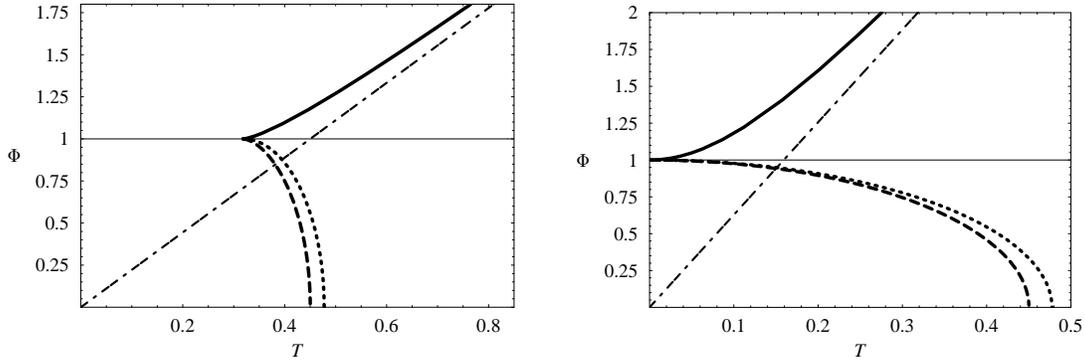}}}
\caption{\footnotesize
  The instability curves of the cases with $(q,0,0)$ and
  $(q,q,q)$ are plotted on the left and right diagrams, respectively.
  The thick solid and dashed curves represent the local stability
  thresholds, dotted curves are the Hawking-Phase phase transition
  lines, the dash-dot lines are the local instability lines of
  the $k=0$ cases and the thin solid lines are the metastability lines.
  The thick solid curve and the dash-dot line merge at large value of $T$.
}\label{fig:D3k1Local}%
}
\end{figure}

The local instability curves are shown in Figure~\ref{fig:D3k1Local}
for some cases, superposed with the instability lines of $k=0$,
the metastability lines and
also with the Hawking-Page phase transition curves.
One of the local instability curves asymptotes to the $k=0$ instability
line.
This is expected, because the larger $T$ is, the larger $r_H$ also is,
and large $k=1$ black hole approaches the $k=0$ case in this limit.
Note also that $k=1$ black hole does not exist for sufficiently
low temperature except for the $(q,q,q)$ configuration.
(See Ref. \cite{Yamada:2007gb} for the interpretation of
the low temperature region.)

\section{M2 and M5 Cases}\label{sec:M2M5}

We have carried out the similar analyses for
the M2 and M5 cases in Appendices \ref{app:M2} and \ref{app:M5}.
As far as the metastability is concerned, we find the same results
as the D3 case, namely, the metastability sets in at $\Phi=0$ and
$1$ for $k=0$ and $1$, respectively.
Though we do not know the corresponding conformal field theories
for those cases, we can expect that they exhibit the similar 
metastabilities as the $\mathcal{N}=4$ theory observed in
Ref.~\cite{Yamada:2006rx}.

As an aside, we briefly mention the low temperature structure
of these $k=1$ cases.
In general, there are two local instability curves and a Hawking-Page
phase transition curve.
Those curves end at the points where the horizon radii go to zero.
The end points are all at $\Phi=1$ but they may not be at zero temperature.
Let us denote the end point of the curve that asymptotes to the $k=0$
line by $T_0$, the one for the other instability curve by $T_1$ and
the one for the Hawking-Page phase transition curve by $T_2$.
For the D3 case in Figure~\ref{fig:D3k1Local}, 
we have $T_0=T_1=T_2=1/\pi$ for $(q,0,0)$ and for $(q,q,q)$,
$T_0=T_1=T_2=0$.
Now for the M2 and M5 cases, those temperatures may not
coincide.
We list those temperatures together with the asymptotic slopes
of one of the instability curves that merge to the $k=0$ instability
straight lines.
See also Figures~\ref{fig:M2Local} and \ref{fig:M5Local} in the appendices.
\begin{table}[h]
  \centerline{
  \begin{tabular}{c|cccc||c|cc}
    M2 & $(q,0,0,0)$ & $(q,q,0,0)$ & $(q,q,q,0)$ & $(q,q,q,q)$ & M5 & $(q,0)$ & $(q,q)$ \\ \hline
 Slope & $\sqrt{2/3}\pi$ & $\pi$ & $4\pi/3$ & $2\pi$ & Slope & $\pi/\sqrt{3}$ & $\pi$ \\
 $T_0$ & $3/(4\pi)$ & $1/(2\pi)$ & $1/(4\pi)$ & $0$  & $T_0$ & $\sqrt{2}/\pi$ & $1/(2\pi)$ \\
 $T_1$ & $1/(\sqrt{2}\pi)$ & $1/(2\pi)$ & $1/(4\pi)$ & $0$  & $T_1$ & $3/(2\pi)$ & $1/(2\pi)$ \\
 $T_2$ & $3/(4\pi)$ & $1/(2\pi)$ & $1/(4\pi)$ & $0$  & $T_2$ & $3/(2\pi)$ & $1/(2\pi)$
  \end{tabular}
  }
\label{table:M2M5}
\end{table}
This is a curious phenomenon, but so far, the physical significance
is unclear.

\section{Conclusion}
\label{sec:conclusion}
We have carried out the probe analysis of the geometry generated by
the spinning branes.
It is found that the probe tends to tunnel through the potential
barrier and fly away to infinity when the chemical potential of
the system ({\it i.e.}, the angular velocity in the five-sphere)
goes beyond the critical value.
The probe can be interpreted as describing the behavior of a brane
splitting from the spinning stack.
Therefore, our finding implies the instability of the spinning
branes by emitting the branes one at a time.
Apart from the fragmentation by tunneling, the probe is locally
stable near the stack at any finite temperature.
This suggests that the fragmentation process analyzed in this work
is different from the known local thermodynamic instability of
the spinning branes.

By analyzing the reduced action of the corresponding black hole
system, we have shown that the instability of the branes manifests
as the metastability of the black hole.
In this analysis, we obtained the metastable critical line as well as
the local stability critical curves.
The metastable critical lines exactly coincide with those of
the probe analysis and they are distinct from the local instability
curves.

We have argued that the fragmentation is due to the violation 
of the BPS bound at the boundary.
For the D3 case, we have the dual field theory
-- the $\mathcal{N}=4$ SYM theory with $R$-symmetry chemical potentials --
and its metastability also is due to the violation of the BPS bound
\cite{Yamada:2006rx}.
The metastability critical line of the field theory phase diagram
precisely coincide with the one found in this paper,
as expected from the independence of the bound on the coupling strength.
Moreover, as explained in the introduction, the runaway behavior of
the split brane nicely corresponds to the similar behavior of
the scalar field eigenvalue in the field theory side.
Thus we have the complete agreement in the AdS/CFT conjecture,
in this regard,
and this work
provides the reason and mechanism of the metastability.
(The analysis of Ref. \cite{Yamada:2006rx} is carried out for
the field theory defined on $S^3$ and corresponds to the $k=1$ case
of this paper.
However, it is clear that when the theory is defined on
$\mathbb{R}^3$, the lack of the conformal curvature coupling of the scalars
makes the critical value of the chemical potential zero,
in agreement with the $k=0$ case of this work.)

We remark that the critical value of the chemical potential for
the $k=1$ case is $\Phi=1/R$ and this is the angular velocity of
the five-sphere whose linear velocity is at the speed of light.
Therefore, for this case, we observe an interesting connection
between the BPS bound and the limit imposed by the speed of light.

As was mentioned in the introduction, 
the brane picture of the local thermodynamic instability is
yet to be seen.
This may involve a more drastic fragmentation into bigger chunks,
if the consideration on the entropy favors such process.
This could also be the localization of the charges as discussed
in Ref. \cite{Cvetic:1999rb}.

\bigskip
\bigskip

\section*{Acknowledgments}
At the beginning of this project,
I benefited from the discussions with Chris Herzog.
I would like to thank the members of the Racah institute 
high energy theory group who provided valuable comments.
I would also like to thank Andy O'Bannon for his useful
suggestions on the manuscript.
This work was supported by the Lady Davis fellowship.

\bigskip
\bigskip

\pagebreak

\appendix

\section{Spinning M2-Branes}\label{app:M2}
In this appendix, we carry out the detailed analysis for spinning
M2-branes, similar to the D3 case.
The discussion closely parallels the main text.

\subsection{Immersing Charged AdS$_4$ Black Holes
in Eleven Dimensions}\label{subsec:M2review}
In Ref.~\cite{Cvetic:1999xp}, the immersion of the charged AdS$_4$
black holes in eleven dimensions is also discussed.
The black hole is a solution to the $\mathcal{N}=2$ gauged supergravity
in four dimensions and the metric is given as \cite{Duff:1999gh,Sabra:1999ux}
\begin{equation}\label{eq:d4metric}
  ds_4^2 = - H^{-1/2} f dt^2
           + H^{1/2} \big( f^{-1}d\rho^2 + \rho^2 d\Omega_{2,k}^2 \big)
  \;,
\end{equation}
where $H:=H_1 H_2 H_3 H_4$, $H_i:=1+q_i/\rho$,
$f=k - \rho_0/\rho + (\rho/R)^2 H$,
$\rho_0 = \rho_H\{k+(\rho_H/R)^2H(\rho_H)\}$, and $\rho_H$ satisfies
$f(\rho_H)=0$.
We again concentrate on the cases with $k=0$ and $1$.
The solutions to the scalar fields $X_i$ and the gauge fields $A_i^{_{(1)}}$
are
\begin{equation}\label{eq:M2others}
  X_i = H^{1/4}/H_i
  \;,\quad
  \text{and}
  \quad
  A_i^{_{(1)}} = - \frac{e_i}{\rho + q_i} dt
  \;,\quad
  \text{with}
  \quad
  e_i := \sqrt{q_i (\rho_0+kq_i)}
  \;.
\end{equation}
We note that the expression for $e_i$ is valid for $k=0$ and $1$, and
not for $k=-1$.

The $S^7$-reduction metric ansatz for the eleven dimensional supergravity is
given as
\begin{equation}\label{eq:11dmetricM2}
  ds_{11}^2 = \tilde\Delta^{2/3} ds_4^2
            + (2R)^2 \tilde\Delta^{-1/3}
            \sum_{i=1}^{4}X_i^{-1} 
            \big\{d\mu_i^2+\mu_i^2\big(d\phi_i+(2R)^{-1}A_i^{_{(1)}}\big)^2\big\}
  \;,
\end{equation}
where $\tilde\Delta := \sum_iX_i\mu_i^2$ and the constraints
$\Pi_iX_i\equiv 1$ and $\sum_i\mu_i^2=1$ are satisfied.
The part $ds_4^2$ is where we would like the metric
(\ref{eq:d4metric}) to fit in as an eleven dimensional solution.
We also have the ansatz for the four-form field strength as
\begin{align}
  F^{_{(4)}} =&
     R^{-1}\sum_i \big( X_i^2\mu_i^2 - \tilde\Delta X_i \big) \epsilon^{_{(4)}}
     - R \sum_i(\bar{*}d\ln X_i)\wedge d(\mu_i^2)
     \nonumber\\
     &+ 2R^2 \sum_iX_i^{-2}d(\mu_i^2)\wedge \big(d\phi_i+(2R)^{-1}A_i^{_{(1)}}\big)
                \wedge \bar{*}F_i^{_{(2)}}
  \;,
\end{align}
where $\epsilon^{_{(4)}}$ is the volume form with respect to the metric
$ds_4^2$,  $\bar{*}$ denotes the Hodge dual also with respect to
the same four dimensional metric and the signs are corrected from
the original.

Plugging these ansatz into the equations of motion of the eleven dimensional
supergravity, one obtains the four dimensional equations of motion.
The equations (\ref{eq:d4metric}) and (\ref{eq:M2others})
satisfy the latter equations of motion
and indeed the four dimensional metric 
fits into the eleven dimensional metric (\ref{eq:11dmetricM2})
as desired.
The three-form field $A^{_{(3)}}$ that satisfies $F^{_{(4)}}=dA^{_{(3)}}$
is
\begin{equation}
  A^{_{(3)}} = \bigg[ \bigg(\frac{\rho}{R}\bigg)^3 \Delta
              - \sum_i \frac{\rho_0+k(-\rho_H+q_i)}{R} \mu_i^2
              \bigg] dt \wedge \epsilon^{_{(2)}}
              +
              2\sum_i\left(\frac{e_i}{R}\right)
                \mu_i^2 (Rd\phi_i)\wedge\epsilon^{_{(2)}}
  \;,
\end{equation}
where $\Delta := H^{3/4}\tilde\Delta$ and the volume form $\epsilon^{_{(2)}}$
is with respect to $R^2d\Omega_{2,k}^2$.
We have corrected a sign in the three-form and included the constant
$(\rho_0-k\rho_H)/R$ which does not contribute to the four-form
$F^{_{(4)}}$ but makes the value of the probe action at $\rho=\rho_H$ zero.
As for the D3 case, Ref.~\cite{Cvetic:1999xp} explicitly demonstrates
that the metric (\ref{eq:11dmetricM2}) with $k=0$ arises
as the near horizon limit of the geometry 
generated by spinning M2-branes.

\subsection{Probing the Geometry of Post-Near-Horizon Limit}
\label{subsec:M2post-NHL}
\subsubsection{Probe Action}
\label{subsubsec:M2probeAction}
We now want to probe the geometry described by the metric
(\ref{eq:11dmetricM2}) with (\ref{eq:d4metric}).
We note that the solution has the inverse temperature
$\beta = 4\pi(H^{1/2}/f')|_{\rho=\rho_H}$ and the angular
velocities of the horizon $\Omega_{H\,i}=-A_{i\,0}(\rho_H)/R$.
We apply the coordinate transformation
\begin{equation}
  \phi_i \to \phi_i + \Omega_{H\,i} \, t
  \;,
\end{equation}
and move onto the co-rotating frame.
This transformation has the only effect in the metric by the shift
\begin{equation}
  A_{i\,0} = \frac{e_i}{\rho_H+q_i} - \frac{e_i}{\rho+q_i}
  \;,
\end{equation}
and we define the chemical potentials of the system by
\begin{equation}
  \Phi_i := A_{i\,0}(\infty)/R
          = \frac{1}{R} \frac{e_i}{\rho_H+q_i}
  \;,
\end{equation}
where we inserted the factor of $R$ for the dimensional reason.
The three-form $A^{_{(3)}}$ now takes the new form
\begin{align}
  A^{_{(3)}} =& \bigg[ \bigg(\frac{\rho}{R}\bigg)^3 \Delta
              + \sum_i \frac{1}{R}
                \bigg\{ \frac{e_i^2}{\rho_H+q_i} 
                       - (\rho_0-k\rho_H+kq_i) \bigg\} \mu_i^2
              \bigg] dt \wedge \epsilon^{_{(2)}}
              \nonumber\\
              &+
              2\sum_i\left(\frac{e_i}{R}\right)
                \mu_i^2 (Rd\phi_i)\wedge\epsilon^{_{(2)}}
  \;.
\end{align}

In the following, we are going to analyze the probe immersed parallel
to the horizon, very similar to the D3 case.
The probe action has the form
\begin{equation}
  V_{M2} = WV - WZ
  \;,
\end{equation}
where we omit the overall minus sign and also drop
the volume factor, {\it i.e.}, the integral
of $dt\wedge\epsilon^{_{(2)}}$, and the tension of the brane.
The first term $WV$ represents the world volume part with respect
to the induced metric $G_{M2}$,
\begin{align}
  WV :=& \sqrt{-\det G_{M2}}
      \nonumber\\
      =& \left(\frac{\rho}{R}\right)^2\tilde\Delta
         \bigg( H^{1/2}f - 
            H \tilde\Delta^{-1} \sum_i X_i^{-1} \mu_i^2 A_{i\,0}^2 \bigg)^{1/2}
  \;,
\end{align}
and the second term $WZ$ represents the three-form coupling
of the probe and is given by the pullback of the three-form
$A^{_{(3)}}$ with respect to the immersion map,
\begin{equation}
  WZ := \left( \frac{\rho}{R} \right)^3 \Delta
        + \sum_i\frac{1}{R}
             \bigg\{ \frac{e_i^2}{\rho_H+q_i} 
                     - (\rho_0-k\rho_H+kq_i) \bigg\} \mu_i^2
  \;.
\end{equation}

\subsubsection{The Case with $k=0$}
\label{subsubsec:probeM2k0}
In this subsection, we are going to briefly summarize the results
for the $k=0$ case.
We present the results for the charge configurations
$(q_1,q_2,q_3,q_4)=(q,0,0,0)$ and $(q,q,q,q)$.
One can also work out the cases with $(q,q,0,0)$ and $(q,q,q,0)$
and finds that they are similar to the behavior of $(q,0,0,0)$
configuration.
In the following analyses, we set $R=1$ and measure the quantities
in the units of $R$.

\bigskip

\noindent{\it The $(q,0,0,0)$ Configuration.}
In this case, the probe action contains one of the three angles
that parametrize the directional cosines $\mu_i$.
It is easy to see that the potential is minimized when the angle is
$\pi/2$ and we concentrate on this special angle.
One can plot the potential with various parameters and find the similar
results to Figure~\ref{fig:D3k0q}.

The behavior of the action at asymptotically large $\rho$ can be
seen in the expansions
\begin{equation}
  WV = \rho^3 - \frac{1}{2} \, \Phi^2 \rho + \mathcal{O}(1)
  \;,\quad
  \text{and}
  \quad
  WZ = \rho^3 + \mathcal{O}(1)
  \;.
\end{equation}
Therefore, we see that for any $\Phi>0$, the probe tends to
fly away if it tunnel through the potential barrier.
When $\Phi=0$, the probe feels no force at the boundary of the
background spacetime due to the exact cancellation between the
pull from the $WV$ part and the repulsion from $WZ$.

We can examine the force acting on the probe near the horizon by
taking the derivative of the potential and expand that around the
horizon:
\begin{equation}
  \frac{\partial V_{M2}}{\partial r}
   = \frac{1}{2}\rho_H^2\sqrt{\rho_H(3\rho_H+2q)/(\rho_H+q)}
     \,
     \frac{1}{\sqrt{\rho-\rho_H}}
     + \mathcal{O}(1)
  \;.
\end{equation}
As in the D3 case, the pull from the world volume part of the probe
action wins out over the repulsion from $WZ$ near the horizon.
We also see that there is no local instability.
This is related to the fact that the temperature of the system,
\begin{equation}
  \beta = \frac{4\pi\sqrt{1+q/\rho_H}}{3\rho_H+2q}
  \;,
\end{equation}
never becomes zero.

\bigskip

\noindent{\it The $(q,q,q,q)$ Configuration.}
In this case, the probe action is independent of the angles that
parametrize the directional cosines $\mu_i$.
The plot appears qualitatively the same as Figure~\ref{fig:D3k0qqq}.
The asymptotic structure is similar to the $(q,0,0,0)$ case and
we again have the instability for any $\Phi>0$.
The derivative of the action near the horizon is
\begin{equation}
  \frac{\partial V_{M2}}{\partial r}
  = \frac{1}{2}\sqrt{(3\rho_H-q)(\rho_H+q)^5/\rho_H}
    \,
    \frac{1}{\sqrt{\rho-\rho_H}}
    + \mathcal{O}(1)
  \;.
\end{equation}
We see that in general the world volume pull dominates near the
horizon, except when $q=3\rho_H$.
At the critical value $q=3\rho_H$, the probe brane is locally
unstable at the horizon and when the value of $q$ exceeds the
critical point, the action becomes ill-defined.
This is related to the fact that the temperature of the system,
\begin{equation}
  \beta = \frac{4\pi\rho_H}{(3\rho_H-q)(\rho_H+q)}
  \;,
\end{equation}
becomes zero at $q=3\rho_H$ and beyond which the system is not defined.

\subsubsection{The Case with $k=1$}
\label{subsubsec:probeM2k1}
We are now going to summarize the results for the $k=1$ case.
We present the results for the charge configurations
$(q_1,q_2,q_3,q_4)=(q,0,0,0)$ and $(q,q,q,q)$.
One can also work out the cases with $(q,q,0,0)$ and $(q,q,q,0)$
and finds that they are similar to the behavior of $(q,0,0,0)$
configuration.

\bigskip

\noindent{\it The $(q,0,0,0)$ Configuration.}
As in the $k=0$ case, the probe action contains one of the three angles
that parametrize the directional cosines $\mu_i$ and we again
set that angle to $\pi/2$.
One can plot the action with various parameters and find the similar
results to Figure~\ref{fig:D3k1q}.

The asymptotic behavior of the action can be seen in the expansions
\begin{equation}
  WV = \rho^3 + \frac{1}{2} (1- \Phi^2) \rho + \mathcal{O}(1)
  \;,\quad
  \text{and}
  \quad
  WZ = \rho^3 + \mathcal{O}(1)
  \;.
\end{equation}
Therefore, we see that for $\Phi>1$, the probe tends to
fly away if it tunnels through the potential barrier.
When $\Phi=1$, the probe feels no force at the boundary of the
background spacetime due to the exact cancellation between the
pull from $WV$ and the repulsion from $WZ$.

The derivative of the potential expanded around the horizon is
\begin{equation}
  \frac{\partial V_{M2}}{\partial r}
   = \frac{1}{2}\rho_H^2\sqrt{(3\rho_H^2+2q\rho_H+1)/(\rho_H+q)}
     \,
     \frac{1}{\sqrt{\rho-\rho_H}}
     + \mathcal{O}(1)
  \;.
\end{equation}
We see that there is no local instability.
This is related to the fact that the temperature of the system,
\begin{equation}
  \beta = \frac{4\pi\sqrt{\rho_H(\rho_H+q)}}{3\rho_H^2+2q\rho_H+1}
  \;,
\end{equation}
never becomes zero.

\bigskip

\noindent{\it The $(q,q,q,q)$ Configuration.}
In this case, the probe action is independent of the angles that
parametrize the directional cosines $\mu_i$.
The plot appears qualitatively the same as Figure~\ref{fig:D3k1qqq}.
The asymptotic structure is similar to the $(q,0,0,0)$ case and
we again have the instability for $\Phi>1$.
The derivative of the action near the horizon is
\begin{equation}
  \frac{\partial V_{M2}}{\partial r}
  = \frac{1}{2}(\rho_H+q)
        \sqrt{(3\rho_H^4+8q\rho_H^3+6q^2\rho_H^2+\rho_H^2-q^4)/\rho_H}
    \,
    \frac{1}{\sqrt{\rho-\rho_H}}
    + \mathcal{O}(1)
  \;.
\end{equation}
We see that in general the world volume pull dominates near the
horizon, except when $q$ takes the critical value $q_c$ that
satisfies the equation
$3\rho_H^4+8q_c\rho_H^3+6q_c^2\rho_H^2+\rho_H^2-q_c^4=0$.
At the critical value , the probe brane is locally
unstable at the horizon and when the value of $q$ exceeds the
critical point, the action becomes ill-defined.
This is related to the fact that the temperature of the system,
\begin{equation}
  \beta = 
    \frac{4\pi\rho_H(\rho_H+q)^2}{3\rho_H^4+8q\rho_H^3+6q^2\rho_H^2+\rho_H^2-q^4}
  \;,
\end{equation}
becomes zero at $q_c$ and beyond which the system is not defined.

\subsection{Analysis of Reduced Action}
\label{subsec:reducedM2Action}
In this section, we show that the previously observed
brane fragmentation corresponds to the metastability
of the four dimensional black hole system by analyzing
the reduced action.
We note that the reduced action in a grand canonical ensemble
always takes the form
\begin{equation}
  I_* = \beta \bigg( E - \sum_i e_i\Phi_i \bigg) - S
  \;,
\end{equation}
where $E$ is the energy of the system, $S$ is the entropy
computed from the surface area of the black hole, $e_i$ are
the charges that satisfy Gauss' law and $\beta$ and $\Phi_i$
are the thermodynamic input data imposed at the boundary.
Therefore, rather than deriving the reduced action from scratch,
we postulate
\begin{equation}
  E := \frac{1}{R^2} \bigg( 2\rho_0 + k \sum_i q_i \bigg)
  \;,
  \quad
  S := 4\pi \left( \frac{\rho_H}{R} \right)^3 H(r_H)^{1/2}
  \;,
\end{equation}
where the parameters $e_i$, $\rho_0$ and the function $H$ are defined
in the previous subsections.
Extremized with respect to the parameters
$\rho_H$ and $q_i$, the reduced action correctly reproduces the on-shell
expressions for $\beta$ and $\Phi_i$.
Moreover, the thermodynamic quantities derived from $I_*$
satisfy the first law.

\begin{figure}[h]
{
\centerline{\scalebox{1.2}{\includegraphics{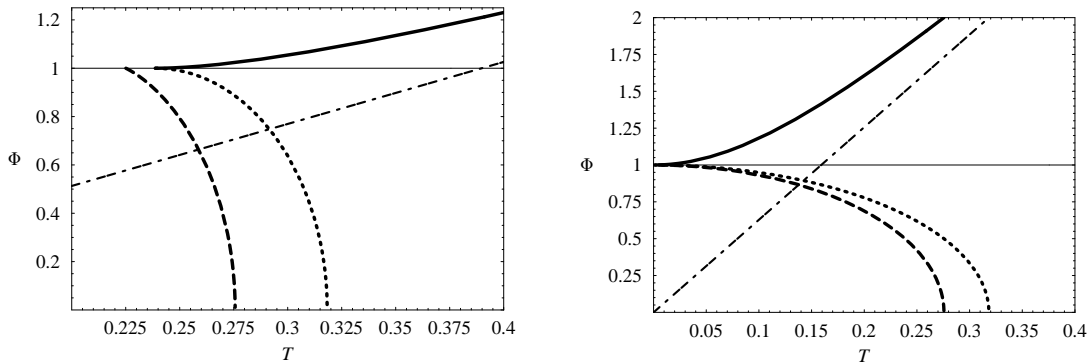}}}
\caption{\footnotesize
  The instability curves of the cases with $(q,0,0,0)$ and
  $(q,q,q,q)$ are plotted on the left and right diagrams, respectively.
  The thick solid and dashed curves represent the local stability
  thresholds, dotted curves are the Hawking-Phase phase transition
  lines, the dash-dot lines are the local instability lines of
  the $k=0$ cases and the thin solid lines are the metastability lines.
  The dash-dot lines both have zero intercepts.
  The thick solid curve and the dash-dot line merge at large value of $T$.
}\label{fig:M2Local}%
}
\end{figure}

The stability of the system can be observed from the behavior of
the action with respect to the parameters $\rho_H$ and $q_i$.
One finds qualitatively the same plots as Figure~\ref{fig:D3Redk0}
for $k=0$ and Figure~\ref{fig:D3Redk1} for $k=1$.
Inspections similar to the D3 case show that the black hole systems
become metastable at the critical values of the chemical potentials
$0$ and $1$ for $k=0$ and $k=1$, respectively.
The local stability can also be investigated from the reduced action.
The results are summarized in Figure~\ref{fig:M2Local}.
As stated in Section~\ref{sec:M2M5},
the local instability curves of the $(q,0,0,0)$ configuration
do not end at the same temperature and the precise values are
listed in Table~\ref{table:M2M5}.
Apart from this, the AdS$_4$ black hole phase diagram has
the similar structure to the five dimensional one in the $T$-$\Phi$
parameter space.
We remark that the phase diagram obtained in Ref.~\cite{Cvetic:1999ne}
is plotted in the space of $\rho_H$ and $q$, and shows a very
different structure to the five dimensional counterpart.

\section{Spinning M5-Branes}\label{app:M5}
In this appendix, we carry out the analysis for spinning
M5-branes, similar to the D3 case.
The discussion closely parallels the main text.

\subsection{Immersing Charged AdS$_7$ Black Holes
in Eleven Dimensions}\label{subsec:M5review}
In Ref.~\cite{Cvetic:1999xp}, the immersion of the charged AdS$_7$
black holes in eleven dimensions is also discussed.
The black hole is a solution to the $\mathcal{N}=4$ gauged supergravity
in seven dimensions and the metric is given as \cite{Cvetic:1999ne}
\begin{equation}\label{eq:d7metric}
  ds_7^2 = - H^{-4/5} f dt^2
           + H^{1/5} \big( f^{-1}d\rho^2 + \rho^2 d\Omega_{5,k}^2 \big)
  \;,
\end{equation}
where $H:=H_1 H_2$, $H_i:=1+q_i/\rho^4$,
$f=k - (\rho_0/\rho)^4 + (\rho/R)^2 H$,
$\rho_0 = \rho_H\{k+(\rho_H/R)^2H(\rho_H)\}^{1/4}$, and $\rho_H$ satisfies
$f(\rho_H)=0$.
We again concentrate on the cases with $k=0$ and $1$.
The solutions to the scalar fields $X_i$ and the gauge fields $A_i^{_{(1)}}$
are
\begin{equation}\label{eq:M5others}
  X_i = H^{2/5}/H_i
  \;,\quad
  \text{and}
  \quad
  A_i^{_{(1)}} = - \frac{e_i}{\rho^4 + q_i} dt
  \;,\quad
  \text{with}
  \quad
  e_i := \sqrt{q_i (\rho_0^4+kq_i)}
  \;.
\end{equation}
We note that the expression for $e_i$ is valid for $k=0$ and $1$, and
not for $k=-1$.

The $S^4$-reduction metric ansatz for eleven dimensional supergravity is
given as
\begin{equation}\label{eq:11dmetricM5}
  ds_{11}^2 = \tilde\Delta^{1/3} ds_7^2
             + (R/2)^2 \tilde\Delta^{-2/3}
               \bigg[ X_0^{-1} d\mu_0^2 +
               \sum_{i=1}^2X_i^{-1} 
               \big\{d\mu_i^2+\mu_i^2\big(d\phi_i+(2/R)A_i^{_{(1)}}\big)^2\big\}
               \bigg]
  \;,
\end{equation}
where $X_0:=(X_1X_2)^{-2}$, $\mu_0^2+\mu_1^2+\mu_2^2=1$ and
$\tilde\Delta := \sum_{\alpha=0}^2X_\alpha\mu_\alpha^2$.
The part $ds_7^2$ is where we would like the metric
(\ref{eq:d7metric}) to fit in as an eleven dimensional solution.
We also have the ansatz for the seven-form field strength as
\begin{align}
  *F^{_{(4)}} =&
      (4/R)\sum_{\alpha=0}^2
      \big( X_\alpha^2\mu_\alpha^2 - \tilde\Delta X_\alpha \big) \epsilon^{_{(7)}}
      +(2/R)\tilde\Delta X_0 \epsilon^{_{(7)}}
      + (R/4)\sum_{\alpha=0}^2 (\bar{*}d\ln X_\alpha)\wedge d(\mu_\alpha^2)
      \nonumber\\
      &+ (R^2/8)\sum_iX_i^{-2}d(\mu_i^2)\wedge \big(d\phi_i+(2/R)A_i^{_{(1)}}\big)
                \wedge \bar{*}F_i^{_{(2)}}
  \;,
\end{align}
where $*$ denotes the Hodge dual in the eleven dimensional metric,
$\epsilon^{_{(7)}}$ is the volume form with respect to the metric
$ds_7^2$,  $\bar{*}$ denotes the Hodge dual also with respect to
the same seven dimensional metric.

Plugging these ansatz into the equations of motion of the eleven dimensional
supergravity, one obtains the seven dimensional equations of motion.
The equations (\ref{eq:d7metric}) and (\ref{eq:M5others})
satisfy the latter equations of motion
and indeed the seven dimensional metric
fits into the eleven dimensional metric (\ref{eq:11dmetricM5})
as desired.
The six-form field $A^{_{(6)}}$ that satisfies $*F^{_{(4)}}=dA^{_{(6)}}$
is
\begin{equation}
  A^{_{(6)}} = \bigg[ \bigg(\frac{\rho}{R}\bigg)^6 \Delta
              - \sum_i \frac{\rho_0^4+k(-\rho_H^4+q_i)}{R^4} \mu_i^2
              \bigg] dt \wedge \epsilon^{_{(5)}}
              +
              \frac{1}{2} \sum_i\left(\frac{e_i}{R^4}\right)
                \mu_i^2 (Rd\phi_i)\wedge\epsilon^{_{(5)}}
  \;,
\end{equation}
where $\Delta := H^{3/5}\tilde\Delta$ and the volume form $\epsilon^{_{(5)}}$
is with respect to $R^2d\Omega_{5,k}^2$.
We have included the constant $(\rho_0^4-k\rho_H^4)/R^4$
which does not contribute to the seven-form
$*F^{_{(4)}}$, but makes the value of the action at $\rho=\rho_H$ zero.
As for the D3 case, Ref.~\cite{Cvetic:1999xp} explicitly demonstrates
that the metric (\ref{eq:11dmetricM5}) with $k=0$ arises
as the near horizon limit of the geometry 
generated by spinning M5-branes.

\subsection{Probing the Geometry of Post-Near-Horizon Limit}
\label{subsec:M5post-NHL}
\subsubsection{Probe Action}
\label{subsubsec:probeM5Action}
We now want to probe the geometry described by the metric
(\ref{eq:11dmetricM5}) with (\ref{eq:d7metric}).
We note that the solution has the inverse temperature
$\beta = 4\pi(H^{1/2}/f')|_{\rho=\rho_H}$ and the angular
velocities of the horizon $\Omega_{H\,i}=-A_{i\,0}(\rho_H)/R$.
We apply the coordinate transformation
\begin{equation}
  \phi_i \to \phi_i + \Omega_{H\,i} \, t
  \;,
\end{equation}
and move onto the co-rotating frame.
This transformation has the only effect in the metric by the shift
\begin{equation}
  A_{i\,0} = \frac{e_i}{\rho_H^4+q_i} - \frac{e_i}{\rho^4+q_i}
  \;,
\end{equation}
and we define the chemical potentials of the system by
\begin{equation}
  \Phi_i := A_{i\,0}(\infty)/R
          = \frac{1}{R} \frac{e_i}{\rho_H^4+q_i}
  \;,
\end{equation}
where we inserted the factor of $R$ for the dimensional reason.
The six-form $A^{_{(6)}}$ now takes the new form
\begin{align}
  A^{_{(6)}} =& \bigg[ \bigg(\frac{\rho}{R}\bigg)^6 \Delta
              + \sum_i \frac{1}{R^4}
                \bigg\{\frac{e_i^2}{\rho_H^4+q_i} 
                       - (\rho_0^4-k\rho_H+kq_i) \bigg\} \mu_i^2
              \bigg] dt \wedge \epsilon^{_{(5)}}
              \nonumber\\
              &+
              \frac{1}{2} \sum_i\left(\frac{e_i}{R^4}\right)
                \mu_i^2 (Rd\phi_i)\wedge\epsilon^{_{(5)}}
  \;,
\end{align}

In the following, we are going to analyze the probe immersed parallel
to the horizon, very similar to the D3 case.
The probe action has the form
\begin{equation}
  V_{M5} = WV - WZ
  \;,
\end{equation}
where we omit the overall minus sign and also drop
the volume factor, {\it i.e.}, the integral
of $dt\wedge\epsilon^{_{(5)}}$, and the tension of the brane.
The first term $WV$ represents the world volume part with respect
to the induced metric $G_{M5}$,
\begin{align}
  WV :=& \sqrt{-\det G_{M5}}
      \nonumber\\
      =& \left(\frac{\rho}{R}\right)^5\tilde\Delta
         \bigg( H^{1/5}f - 
            H \tilde\Delta^{-1} \sum_i X_i^{-1} \mu_i^2 A_{i\,0}^2 \bigg)^{1/2}
  \;,
\end{align}
and the second term $WZ$ represents the six-form coupling
of the probe and is given by the pullback of the six-form
$A^{_{(6)}}$ with respect to the immersion map
\begin{equation}
  WZ := \left( \frac{\rho}{R} \right)^6 \Delta
        + \sum_i\frac{1}{R^4}
             \bigg\{ \frac{e_i^2}{\rho_H^4+q_i} 
                     - (\rho_0^4+kq_i) \bigg\} \mu_i^2
  \;.
\end{equation}

\subsubsection{The Case with $k=0$}
\label{subsubsec:probeM5k0}
In this subsection, we are going to briefly summarize the results
for the $k=0$ case.
We present the results for the charge configurations
$(q_1,q_2)=(q,0)$ and $(q,q)$.
In the following analyses, we set $R=1$ and measure the quantities
in the units of $R$.

\bigskip

\noindent{\it The $(q,0)$ Configuration.}
Let us parametrize the directional cosines as
\begin{equation}\label{eq:direcCos}
  \mu_0 = \sin\theta
  \;,\quad
  \mu_1 = \cos\theta \sin\psi
  \;,\quad
  \text{and}
  \quad
  \mu_2 = \cos\theta \cos\psi
  \;.
\end{equation}
The action in this case contains both $\theta$ and $\psi$.
It is easy to see that the potential is minimized when the angles are
set to $\theta=0$ and $\psi=\pi/2$
and we concentrate on this special angles.
One can plot the potential with various parameters and find the similar
results to Figure~\ref{fig:D3k0q}.

The behavior of the action at asymptotically large $\rho$ can be
seen in the expansions
\begin{equation}
  WV = \rho^6 - \frac{1}{2} \Phi^2 \rho^4
        - \frac{1}{8}  \Phi^4 \rho^2 + \mathcal{O}(1)
  \;,\quad
  \text{and}
  \quad
  WZ = \rho^6 + \mathcal{O}(1)
  \;.
\end{equation}
Therefore, we see that for any $\Phi>0$, the probe tends to
fly away if it tunnels through the potential barrier.
When $\Phi=0$, the probe feels no force at the boundary of the
background spacetime due to the exact cancellation between the
pull from $WV$ and the repulsion from $WZ$.

We can examine the force acting on the probe near the horizon by
taking the derivative of the potential and expand that around the
horizon:
\begin{equation}
  \frac{\partial V_{M5}}{\partial r}
   = \rho_H^5\sqrt{\rho_H(3\rho_H^4+q)/2(\rho_H^4+q)}
     \,
     \frac{1}{\sqrt{\rho-\rho_H}}
     + \mathcal{O}(1)
  \;.
\end{equation}
As in the D3 case, the pull from the world volume part of the probe
action wins out over the repulsion from $WZ$ near the horizon.
We also see that there is no local instability.
This is related to the fact that the temperature of the system,
\begin{equation}
  \beta = \frac{2\pi\rho_H\sqrt{\rho_H^4+q}}{3\rho_H^4+q}
  \;,
\end{equation}
never becomes zero.

\bigskip

\noindent{\it The $(q,q)$ Configuration.}
In this case, the probe action depends on the angle $\theta$.
One can easily verify that the potential is minimized when $\theta=0$
and we concentrate on this special angle.
The plot appears qualitatively the same as Figure~\ref{fig:D3k0qqq}.
The asymptotic structure is similar to the $(q,0)$ case and
we again have the instability for any $\Phi>0$.
The derivative of the potential near the horizon is
\begin{equation}
  \frac{\partial V_{M5}}{\partial r}
  = \rho_H\sqrt{\rho_H(3\rho_H^4-q)(\rho_H^4+q)/2}
    \,
    \frac{1}{\sqrt{\rho-\rho_H}}
    + \mathcal{O}(1)
  \;.
\end{equation}
We see that in general the world volume pull dominates near the
horizon, except when $q=3\rho_H^4$.
At the critical value $q=3\rho_H^4$, the probe brane is locally
unstable at the horizon and when the value of $q$ exceeds the
critical point, the action becomes ill-defined.
This is related to the fact that the temperature of the system,
\begin{equation}
  \beta = \frac{2\pi\rho_H^3}{3\rho_H^4-q}
  \;,
\end{equation}
becomes zero at $q=3\rho_H^4$ and beyond which the system is not defined.

\subsubsection{The Case with $k=1$}
\label{subsubsec:probeM5k1}
We are now going to summarize the results for the $k=1$ case.
We present the results for the charge configurations
$(q_1,q_2)=(q,0)$ and $(q,q)$.

\bigskip

\noindent{\it The $(q,0)$ Configuration.}
As in the $k=0$ case, the probe action depends on the angles $\theta$
and $\psi$ defined in (\ref{eq:direcCos}).
The potential is minimized when $\theta=0$ and $\psi=\pi/2$ and we
concentrate on these special angles.
One can plot the action with various parameters and find the similar
results to Figure~\ref{fig:D3k1q}.

The asymptotic behavior of the action can be seen in the expansions
\begin{equation}
  WV = \rho^6 + \frac{1}{2} (1- \Phi^2) \rho^4
        - \frac{1}{8} (1- \Phi^2)^2 \rho^2 + \mathcal{O}(1)
  \;,\quad
  \text{and}
  \quad
  WZ = \rho^6 + \mathcal{O}(1)
  \;.
\end{equation}
Therefore, we see that for $\Phi>1$, the probe tends to
fly away if it tunnels through the potential barrier.
When $\Phi=1$, the probe feels no force at the boundary of the
background spacetime due to the exact cancellation between the
pull from $WV$ and the repulsion from $WZ$.

The derivative of the potential expanded around the horizon is
\begin{equation}
  \frac{\partial V_{M5}}{\partial r}
   = \rho_H^5\sqrt{\rho_H(3\rho_H^4+2\rho_H^2+q)/2(\rho_H+q)}
     \,
     \frac{1}{\sqrt{\rho-\rho_H}}
     + \mathcal{O}(1)
  \;.
\end{equation}
We see that there is no local instability.
This is related to the fact that the temperature of the system,
\begin{equation}
  \beta = \frac{2\pi\rho_H\sqrt{\rho_H^4+q}}{3\rho_H^4+2\rho_H^2+q}
  \;,
\end{equation}
never becomes zero.

\bigskip

\noindent{\it The $(q,q)$ Configuration.}
In this case, the probe action depends on $\theta$.
The potential is minimized when $\theta=0$ and we concentrate on this angle.
The plot appears qualitatively the same as Figure~\ref{fig:D3k1qqq}.
The asymptotic structure is similar to the $(q,0)$ case and
we again have the instability for $\Phi>1$.
The derivative of the potential near the horizon is
\begin{equation}
  \frac{\partial V_{M5}}{\partial r}
  = \rho_H\sqrt{\rho_H(3\rho_H^8+2\rho_H^6+2q\rho_H^4-q^2)/2}
    \,
    \frac{1}{\sqrt{\rho-\rho_H}}
    + \mathcal{O}(1)
  \;.
\end{equation}
We see that in general the world volume pull dominates near the
horizon, except when $q$ takes the critical value $q_c$ that
satisfies the equation
$3\rho_H^8+2\rho_H^6+2q_c\rho_H^4-q_c^2=0$.
At the critical value , the probe brane is locally
unstable at the horizon and when the value of $q$ exceeds the
critical point, the action becomes ill-defined.
This is related to the fact that the temperature of the system,
\begin{equation}
  \beta = 
    \frac{2\pi\rho_H^3(\rho_H^4+q)}{3\rho_H^8+2\rho_H^6+2q\rho_H^4-q^2}
  \;,
\end{equation}
becomes zero at $q_c$ and beyond which the system is not defined.

\subsection{Analysis of Reduced Action}
\label{subsec:reducedM5Action}
In this section, we show that the previously observed
brane fragmentation corresponds to the metastability
of the seven dimensional black hole system by analyzing
the reduced action.
As for the M2 case, we postulate
\begin{equation}
  I_* = \beta \bigg( E - \sum_i e_i\Phi_i \bigg) - S
  \;,
\end{equation}
where
\begin{equation}
  E := \frac{1}{R^5} \bigg( \frac{5}{4}\rho_0^4 + k \sum_i q_i \bigg)
  \;,
  \quad
  S := \pi \left( \frac{\rho_H}{R} \right)^5 H(r_H)^{1/2}
  \;,
\end{equation}
where the parameters $e_i$, $\rho_0$ and the function $H$ are defined
in the previous subsections.
Extremized with respect to the parameters
$\rho_H$ and $q_i$, the reduced action correctly reproduces the on-shell
expressions for $\beta$ and $\Phi_i$.
Moreover, the thermodynamic quantities derived from $I_*$
satisfy the first law.

\begin{figure}[h]
{
\centerline{\scalebox{1.2}{\includegraphics{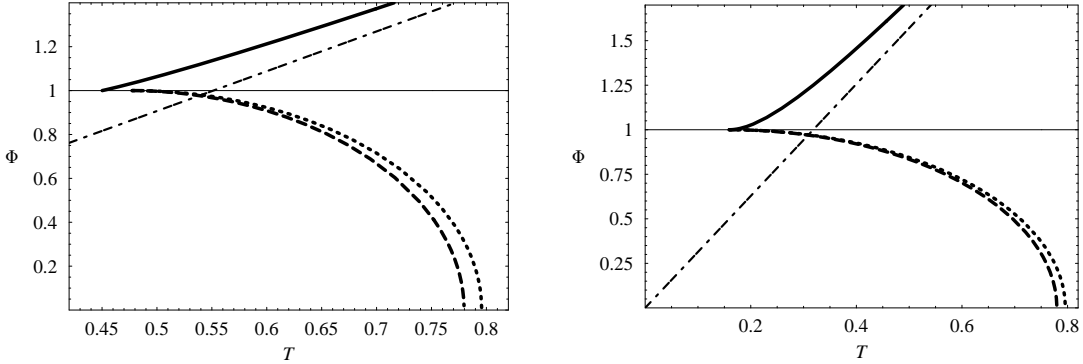}}}
\caption{\footnotesize
  The instability curves of the cases with $(q,0)$ and
  $(q,q)$ are plotted on the left and right diagrams, respectively.
  The thick solid and dashed curves represent the local stability
  thresholds, dotted curves are the Hawking-Phase phase transition
  lines, the dash-dot lines are the local instability lines of
  the $k=0$ cases and the thin solid lines are the metastability lines.
  The dash-dot lines both have zero intercepts.
  The thick solid curve and the dash-dot line merge at large value of $T$.
}\label{fig:M5Local}%
}
\end{figure}

The stability of the system can be observed from the behavior of
the action with respect to the parameters $\rho_H$ and $q_i$.
One finds qualitatively the same plots as Figure~\ref{fig:D3Redk0}
for $k=0$ and Figure~\ref{fig:D3Redk1} for $k=1$.
Inspections similar to the D3 case show that the black hole systems
become metastable at the critical values of the chemical potentials
$0$ and $1$ for $k=0$ and $k=1$, respectively.
The local stability can also be investigated from the reduced action.
The results are summarized in Figure~\ref{fig:M5Local}.
As stated in Section~\ref{sec:M2M5},
the local instability curves of the $(q,0)$ configuration
do not end at the same temperature and the precise values
are listed in Table~\ref{table:M2M5}.
Apart from this, the AdS$_7$ black hole phase diagram has
the similar structure to the five dimensional one in the $T$-$\Phi$
parameter space.
We remark that the phase diagram obtained in Ref.~\cite{Cvetic:1999ne}
is plotted in the space of $\rho_H$ and $q$, and shows a very
different structure to the five dimensional counterpart.

\bigskip


\end{document}